\documentclass[apj]{emulateapj}

\usepackage{times}
\usepackage{epstopdf}
\usepackage{subfigure}
\usepackage{amsmath}

\renewcommand{\ion}[2]{#1\,{\textsc{\romannumeral#2}}}

\newcommand{\irisLong}{\textit{Interface Region Imaging Spectrograph}}
\newcommand{\iris}{\textit{IRIS}}
\newcommand{\rhessiLong}{\textit{The Reuven Ramaty High Energy Solar Spectroscopic Imager}}
\newcommand{\rhessi}{\textit{RHESSI}}
\newcommand{\xrt}{\textit{XRT}}
\newcommand{\xrtLong}{\textit{X-Ray Telescope}}
\newcommand{\eis}{\textit{EIS}}
\newcommand{\eisLong}{\textit{Extreme Ultraviolet Imaging Spectrometer}}
\newcommand{\aia}{\textit{AIA}}
\newcommand{\aiaLong}{\textit{Atmospheric Imaging Assembly}}
\newcommand{\yohkoh}{\textit{Yohkoh}}
\newcommand{\trace}{\textit{TRACE}}
\newcommand{\goes}{\textit{GOES}}
\newcommand{\foxsi}{\textit{FOXSI}}
\newcommand{\hinode}{\textit{Hinode}}
\newcommand{\sdo}{\textit{SDO}}

\makeatletter
\newcommand*{\rom}[1]{\expandafter\@slowromancap\romannumeral #1@}
\makeatother

\shorttitle{Signatures of Impulsive Heating Events. II. Modeling}
\shortauthors{Reep et al.}

\begin{document}


\title{Transition Region and Chromospheric Signatures of Impulsive Heating Events. II. Modeling}

\author{Jeffrey W. Reep\altaffilmark{2,1}, Harry P. Warren\altaffilmark{1}, Nicholas
  A. Crump\altaffilmark{3}, Paulo J. A. Sim\~{o}es\altaffilmark{4}}

\altaffiltext{1}{Space Science Division, Naval Research Laboratory, Washington, DC 20375}
\altaffiltext{2}{National Research Council Postdoctoral Fellow}
\altaffiltext{3}{Naval Center for Space Technology, Naval Research Laboratory, Washington, DC 20375}
\altaffiltext{4}{SUPA School of Physics and Astronomy, University of Glasgow, Glasgow G12 8QQ, UK}


\begin{abstract}
Results from the Solar Maximum Mission showed a close connection between the hard X-ray and
transition region emission in solar flares.  Analogously, the modern combination of \rhessi\ and
\iris\ data can inform the details of heating processes in ways never before possible.  We study a
small event that was observed with \rhessi, \iris, \sdo, and \hinode, allowing us to strongly
constrain the heating and hydrodynamical properties of the flare, with detailed observations
presented in a previous paper.  Long duration red-shifts of transition region lines observed in
this event, as well as many other events, are fundamentally incompatible with
chromospheric condensation on a single loop.  We combine \rhessi\ and \iris\ data to measure the
energy partition among the many magnetic strands that comprise the flare.  Using that
observationally determined energy partition, we show that a proper multi-threaded model can
reproduce these red-shifts in magnitude, duration, and line intensity, while simultaneously being
well constrained by the observed density, temperature, and emission measure.  We comment on the
implications for both \rhessi\ and \iris\ observations of flares in general, namely that: (1) a
single loop model is inconsistent with long duration red-shifts, among other observables; (2) the
average time between energization of strands is less than 10 seconds, which implies that for a hard
X-ray burst lasting ten minutes, there were at least 60 strands within a single \iris\ pixel
located on the flare ribbon; (3) the majority of these strands were explosively heated with energy
distribution well described by a power law of slope $\approx -1.6$; (4) the multi-stranded model
reproduces the observed line profiles, peak temperatures, differential emission measure
distributions, and densities.
\end{abstract}



\section{introduction}
\label{sec:intro}

The transport of energy through flaring coronal loops is well studied, both observationally and theoretically, but not yet fully understood.  The release of energy from magnetic reconnection events drives the acceleration of particles, generation of waves, and {\it in situ} heating of the coronal plasma, although it is not clear how energy is partitioned between the mechanisms.  Further complicating the problem is that the partition of energy amongst the loops that comprise the arcade that forms along the flare ribbon has not been determined to date.  

Flare energy release undoubtedly occurs across many magnetic threads, as has been known for a long time ({\it e.g.} \citealt{svestka1982}).  \citet{aschwanden2001} presented an analysis of a large flare occurring across more than 100 loops, to infer cooling times across a well-observed arcade.  \yohkoh\ observations pointed to a temperature gradient in the arcade, where the outermost loops are the hottest \citep{tsuneta1996}.  Tracing the motion of hard X-ray (HXR) sources, \citet{grigis2005} showed that as a disturbance propagates along the arcade, it triggers reconnection and particle acceleration in successive loops as it proceeds, thus heating the loops sequentially.  

Multi-threaded models have been employed by a number of authors to study solar flares.  \citet{hori1997,hori1998} adopted a multi-stranded model to explain the observation of stationary \ion{Ca}{19} emission during the impulsive phase of many flares, when single loop models consistently predicted strong blue-shifts.  Similarly, \citet{reeves2002} developed a multi-threaded model to show that \trace\ and \yohkoh\ light-curves were more readily explained by an arcade rather than a single loop.  \citet{warren2005} derived an algorithm to compute energy inputs for successive threads comprising a flare, by calculating the discrepancy between the observed and calculated GOES flux.  They showed that the absence of strongly blue-shifted \ion{Ca}{19} emission in \yohkoh\ observations is because that emission is masked by previously heated threads.  \citet{warren2006} studied the duration of heating on successive threads, concluding that short heating time scales lead to significantly higher temperatures, inconsistent with \yohkoh\ observations.  \citet{falewicz2015} compared one and two-dimensional models of a flare to find that the observed dynamics were better reproduced by their 2D model, which approximated a multi-stranded model.  On the other hand, \citet{doschek2015b} found that while a single loop model can reproduce high temperature evaporation flows, there were numerous discrepancies between the observed and modeled cooler, red-shifted lines.  Recently, \citet{qiu2016}, using the 0D model EBTEL \citep{klimchuk2008}, studied the cooling phase of flares with a multi-threaded model, and only found consistency with EUV emission if there is prolonged gradual phase heating occurring on many threads.   

In the first paper (\citealt{warren2016}, hereafter Paper \rom{1}), we studied extensively a small flare that was seen with \irisLong\ (\iris, \citealt{depontieu2014}), \rhessiLong\ (\rhessi, \citealt{lin2002}), \aiaLong\ (\aia, \citealt{lemen2012}), and \eisLong\ and \xrtLong\ aboard \hinode\ (\eis\ and \xrt, respectively, \citealt{culhane2007} and \citealt{golub2007}).  The combination of instruments allows coverage across a wide temperature range, from the chromosphere through the transition region (TR) and upper corona, to temperatures exceeding 10 MK.  The unique perspective allowed us to measure temperatures, emission measures (EMs), non-thermal electron beam parameters, energy input, and individual TR brightenings at high cadence and spatial resolution.  

In Paper \rom{1}, we presented observations of \ion{Si}{4} and \ion{C}{2} as seen by \iris, both of which brightened during the rise phase along with the HXR emission measured with \rhessi.  The two lines were red-shifted during that time period, and remained red-shifted even after the impulsive phase, gradually decreasing in magnitude over time-scales exceeding 20 minutes at some positions. Similar trends in \ion{Si}{4} and other cool lines were reported by other authors in larger flares seen with \iris, {\it e.g.} \citet{sadykov2015,brannon2015, polito2016}.

In this paper, we seek to explain the persistent red-shifts by developing a model which requires a partition of energy amongst the magnetic strands comprising the flare.  In Section \ref{sec:modeling}, we describe the hydrodynamic code used to model this flare.  We then split up the results in Section \ref{sec:results} into two parts: a simple model (both single loop and multi-threaded loop) and a multi-threaded Monte Carlo simulation.  We finally discuss the implications and conclusions of this work in Section \ref{sec:implications}.

\section{Modeling}
\label{sec:modeling}

We have run hydrodynamic simulations with the HYDrodynamics and RADiation code (HYDRAD; \citealt{bradshaw2003,bradshaw2013}) in order to study potential heating mechanisms.  The code solves the one-dimensional hydrodynamic equations, conservation of mass, momentum, and energy, applicable to a two-fluid plasma confined to a magnetic flux tube rooted beneath the surface and extending into the corona.  The energy equations include terms for thermal conduction, enthalpy flux, small-scale electric fields, viscosity, gravity, inter-species collisions, and radiation.  A key strength of the HYDRAD code is its speed and portability.  In this paper, we present a total of 45 simulations, each for 1,000 seconds of simulation time with highly resolved grids and a wide parameter space.  All of the simulations were performed on a desktop computer, with up to 8 running simultaneously.

Because the corona is a low $\beta$ plasma \citep{reidy1968,dulk1978,gary2001}, cross-field conduction is negligible \citep{goedbloed2004}, and the fields are frozen-in \citep{alfven1943}, each coronal loop may be treated as an isolated structure, with no interaction between adjacent loops.  We assume each thread is semi-circular in shape, oriented vertically from the solar surface, with constant cross-section from foot-point to apex.   

We treat radiative losses in the corona and TR with a full calculation of losses with CHIANTI v.8 \citep{dere1997,delzanna2015}, via the equation ({\it e.g.} \citealt{mason1994, bradshaw2013b}):
\begin{align}
E_{R}(X) &= n^{2} \Lambda \nonumber \\
	&= n^2 \Bigg(0.83 \times Ab(X) \times \sum_{i=1}^{i=Z+1} \epsilon_{i}\ X_{i} \Bigg)
\label{eqn:radlosses}	
\end{align}
\noindent where $n$ is the number density, $Ab(X)$ the abundance of element $X$ relative to hydrogen, $\epsilon_{i}$ the emissivity of all lines from ion $i$ of element $X$, and $X_{i}$ the population fraction of ion $i$.  We solve a continuity equation for non-equilibrium ionization states of hydrogen, carbon, oxygen, silicon, and iron in the work here \citep{bradshaw2003}, as non-equilibrium ionization is expected to be significant during impulsive bursts of heating.  Further, because hydrogen may not be ionized in the chromosphere, collisions and thermal conduction due to neutrals are included in the code \citep{orrall1961}.  We treat optically thick radiation in the chromosphere with the recipes of \citet{carlsson2012}.  We adopt the photospheric abundances of \citet{asplund2005} as some recent evidence indicates that flares are photospheric in composition \citep{warren2014}, although there is also evidence to the contrary \citep{dennis2015,doschek2015}.

We have assumed that the loop is heated by an electron beam under the collisional thick-target model \citep{brown1971,hudson1972}, where accelerated electrons stream through the corona, depositing their energy through collisions with chromospheric plasma.  We have treated heating by an electron beam using the model of \citet{emslie1978}, with the details of the implementation described in \citet{reep2013}.  We assume an electron beam distribution of the form:
\begin{equation}
\mathfrak{F}(E_{0}, t) = \frac{F_{0}(t)}{E_{c}^{2}}\ (\delta - 2) \times
  \begin{cases}
   0 & \text{if } E_{0} < E_{c} \\
   \Big(\frac{E_{0}}{E_{c}}\Big)^{- \delta}       & \text{if } E_{0} \geq E_{c}
  \end{cases}
\label{eqn:sharpdist}
\end{equation}
\noindent where $F_{0}(t)$ is the beam energy flux (erg\,sec$^{-1}$\,cm$^{-2}$), $E_{c}$ is the low energy cut-off (keV), $\delta$ is the spectral index, and $E_{0}$ is the initial kinetic energy of a given electron (keV).  This distribution, referred to as a sharp cut-off, is commonly assumed and allows for easy comparison to measured \rhessi\ data.  We use the actual \rhessi\ parameters in the model when available, although in Section \ref{subsec:montecarlo} we treat the energy input for a given loop as randomly selected on a power-law distribution, as described in that section.

\begin{figure*}
\centering
\includegraphics[width=7.in]{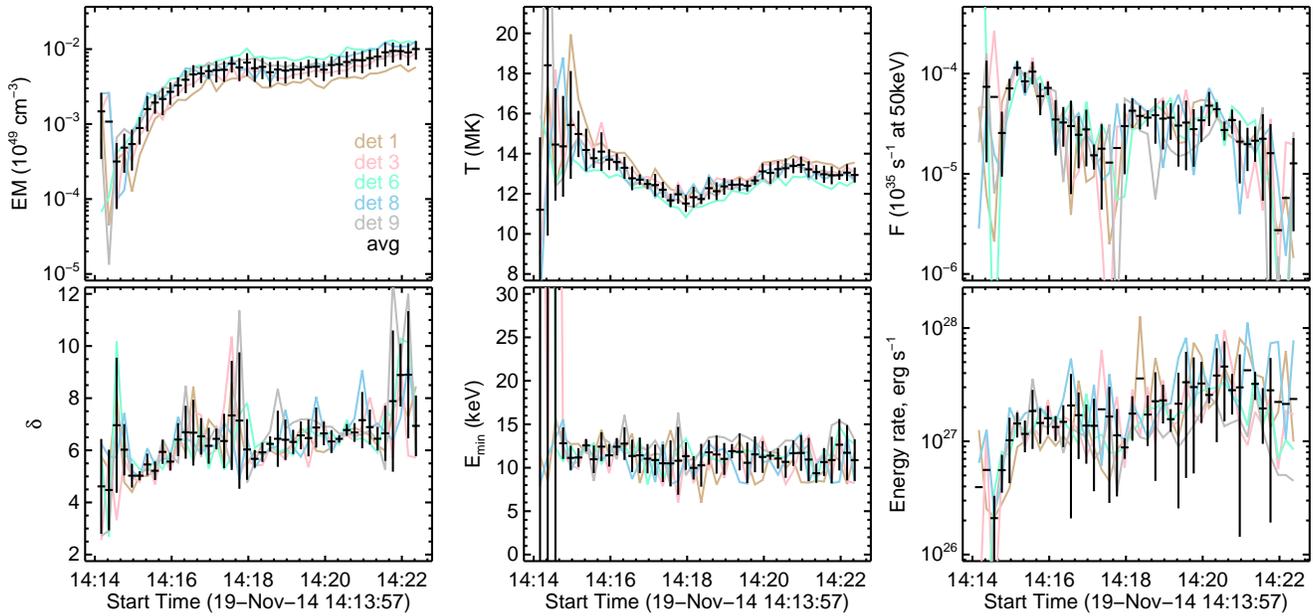}
\caption{The electron beam parameters, as well as temperature and emission measure, as measured with \rhessi\ over the course of the event.  The points in color refer to measurements with a single detector, while the black points denote the average of all detectors.  }
\label{fig:rhessi_params}
\end{figure*}

With \rhessi, although the spectra are not spatially resolved, we measured the electron beam parameters along with temperature and emission measure as a function of time early in the event, shown in Figure \ref{fig:rhessi_params}.  The points in color refer to measurements from a single detector (specifically, detectors 1, 3, 6, 8, and 9), while the black points refer to the mean of all the detectors.  The power carried by the electron beam gradually increases, though it stays around $1 - 3 \times 10^{27}$\,erg\,s$^{-1}$.  A cross-sectional area $A$ must be measured (or assumed) to determine the energy flux $F_{0}(t) = P(t) / A(t)$.  The spectral index $\delta$ gradually increases, but is approximately $6$ at most times (slightly lower than the median value for a microflare, \citealt{hannah2008}).  The cut-off energy $E_{c}$ is approximately $11$\,keV for the entire duration (compared to a median of $12$\,keV in microflares, \citealt{hannah2008}).  

From these simulations, we forward model spectral lines as might be seen by \iris, following the methodology of \citet{bradshaw2011}, using the \iris\ response functions calculated with SolarSoft.  In this work we focus on the \ion{Si}{4} 1402.770 \AA\ and \ion{C}{2} 1334.535 \AA\ lines, which are useful diagnostics of heat transport to the lower atmosphere ({\it e.g.} \citealt{testa2014}), and strongly correlated with energy input \citep{cheng1981}.  

\section{Results}
\label{sec:results}

In Paper \rom{1}, we presented detailed observations of SOL2014-11-19T14:25UT, a small flare which was observed simultaneously by \iris, \rhessi, and \hinode.  The \ion{Si}{4} emission along the slit shows distinct brightenings (typically lasting less than 60\,s) between 14:15 and 14:25, at which times the line profile is red-shifted to $\approx 30$\,km\,s$^{-1}$.  After these brightenings, however, the red-shifts persist, remaining greater than $20$\,km s$^{-1}$ for more than 30 minutes, gradually declining.  Emission from \ion{C}{2} is similar (see Paper \rom{1}).  In this section, we focus on reproducing long-lasted red-shifts, while simultaneously remaining within the constraints given by the other instruments (temperature, density, EM, power).  

\subsection{Simple Model}
\label{subsec:simple}

\begin{figure*}
  \begin{minipage}[t]{0.5\linewidth}
    \centering
    \includegraphics[width=3.3in]{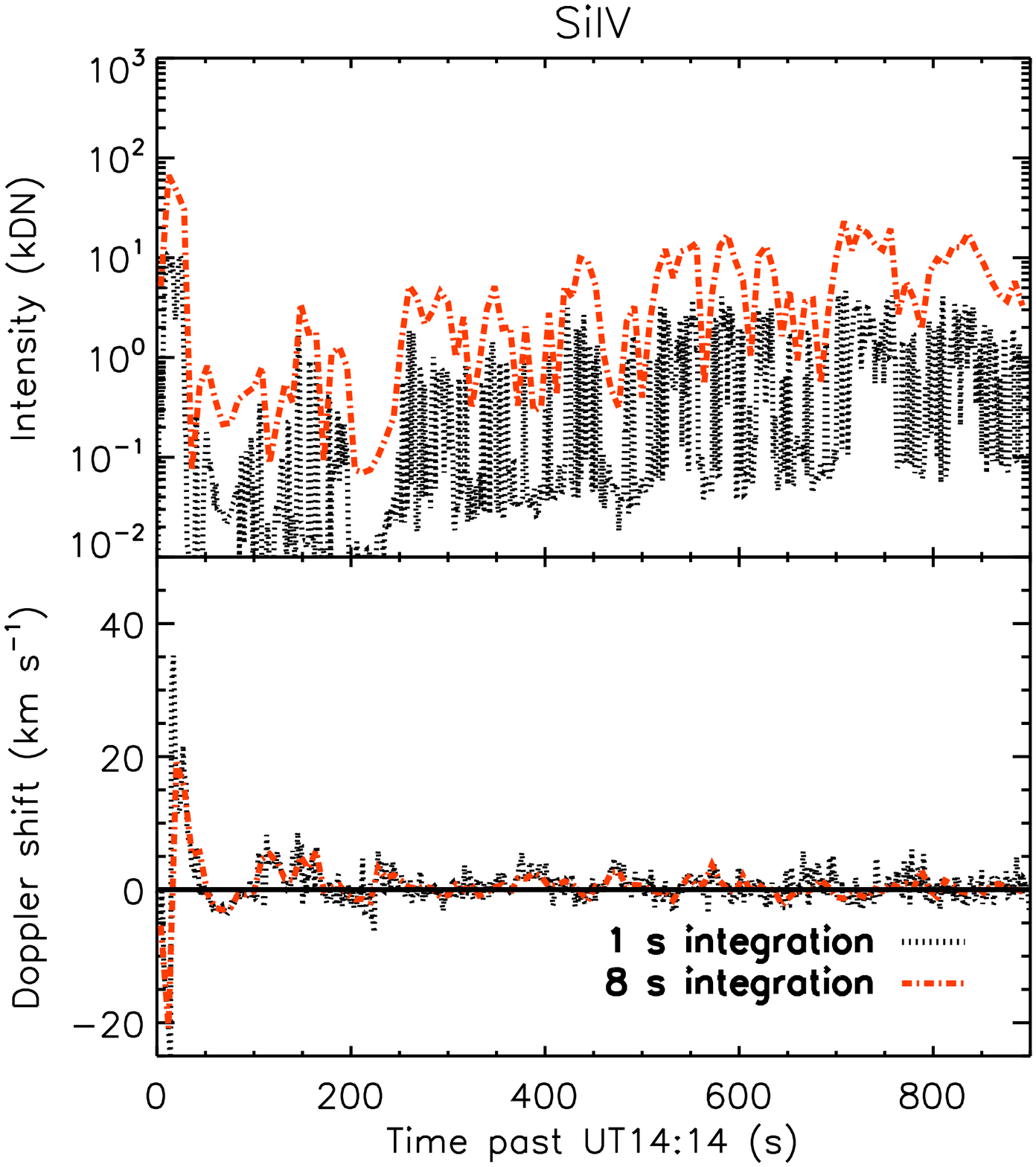}
  \end{minipage}
  \begin{minipage}[b]{0.5\linewidth}
    \centering
    \includegraphics[width=3.3in]{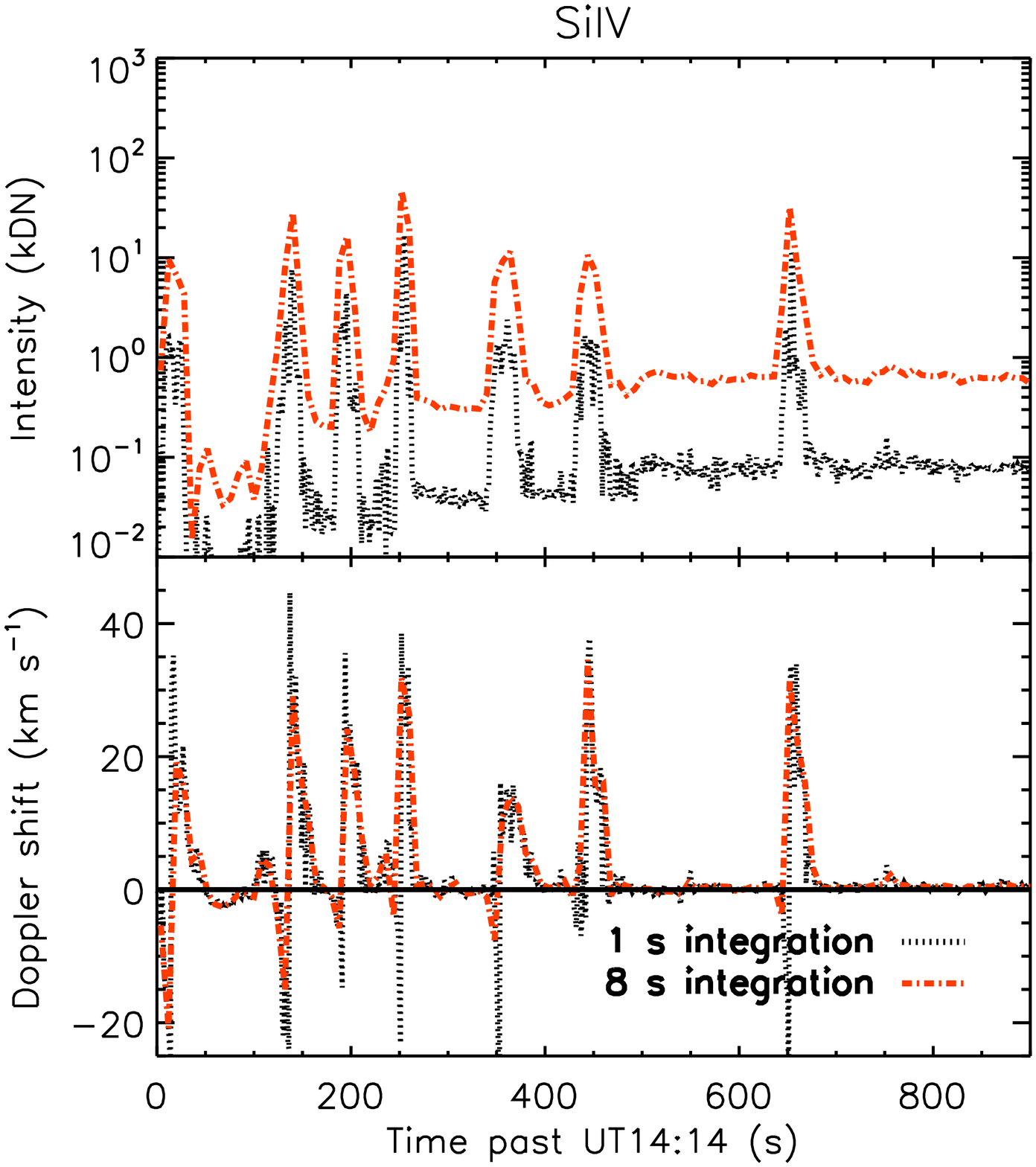}
  \end{minipage}
  \caption{Synthesized \ion{Si}{4} emission relative to \iris\ for the two simulations: seven bursts of heating on a single loop (left) and seven threads with one heating burst each (right).  The top plots show the synthesized intensity as a function of time while the bottom plots show the Doppler shift as a function of time.}
  \label{fig:simulated_Si4}
\end{figure*}

We first attempted to model this event with a single loop, on which seven heating bursts occurred (based on the number of observed \ion{Si}{4} brightenings during the peak intensity).  We also attempted to model this event using a multi-threaded loop composed of seven strands, on each of which one heating burst occurred.  In both cases, we use the observationally measured cut-off energy $E_{c} \approx 11$\,keV and spectral index $\delta \approx 6$, with a maximum beam flux of $5 \times 10^{9}$\,erg\,s$^{-1}$\,cm$^{-2}$ (above the explosive threshold for this cut-off, \citealt{reep2015}).  Figure \ref{fig:simulated_Si4} shows synthesized \ion{Si}{4} emission for the two cases.  On the left hand side, the plots show a single loop modeled with seven distinct bursts of heating, while the right hand shows a simulation with seven threads each with one heating burst.  The top plots show the \ion{Si}{4} intensity as a function of time for 15 minutes while the bottom plots show the Doppler shifts of the line (fit with a single Gaussian).

There is an obvious discrepancy between these simulations and the observations.  First, the intensity in the single thread case is more or less constant after the initial heating event, and there are distinct intensity bursts at each heating burst in the multi-threaded case.  In the observations, however, the \ion{Si}{4} intensity sharply brightens during the rise phase, with a few local maxima, and then very gradually decays.  Second, in the monolithic loop simulation, the Doppler velocity shows only a single strong red-shift at the time of the first heating burst, while the multi-threaded simulation shows seven strong red-shifts, corresponding to a heating event on each thread (which may be more akin to the red-shifts seen by \citealt{brosius2015}).  The observations, however, show a nearly constant red-shift of $> 20$\,km\,s$^{-1}$, with weak spikes in the velocity when the intensity spikes, which neither simulation reproduces.

The quick decay of red-shifts in the simulations is straight-forward to explain.  The heating burst quickly heats the chromosphere to coronal temperatures, which then drives energy down through the transition region, forcing a flow of material downward.  As the material proceeds to greater depths, the local density increases, causing a loss of momentum and a quick dissipation of the bulk flows.  Further, in the monolithic simulation, once evaporation brings material into the corona, later heating bursts cannot cause a strong velocity flow due to the increased inertia.  In the multi-threaded simulation, each loop shows a strong red-shift at the time of a heating burst, as material is pushed downward.  In both cases, the down-flows quickly dissipate in the span of $\approx$20--30\,s.  \citet{fisher1989} analytically derived this quick decay of condensation flows, showing that, at most, they last for 45--60\,s, which is independent of the heating duration (although that does not imply that emission in the same spectral line lasts that long, so it is effectively an upper limit).  

The observed values are irreconcilable with these simulations.  A persistent down-flow of $> 20$\,km\,s$^{-1}$ cannot be explained by constant heating, or with any number of heating bursts on a single loop.  A multi-threaded model could be consistent if the number of threads were significantly higher ($> 50$) than what we have assumed, as we will show in the next section.

\subsection{Monte Carlo Model}
\label{subsec:montecarlo}

The inadequacy of a simple model highlights the complexity of the event.  Neither a single loop model nor a simple multi-threaded model with a small number of threads produces emission consistent with the observed values.  Thanks to the wealth of observations available during this event, we can constrain the parameters of the model more thoroughly and produce a more realistic model.  The first major consideration is that there must have been a very large number of strands comprising the emission {\it within a single \iris\ pixel}.  

The \rhessi\ observations indicate the presence of an electron beam with mean values of low energy cut-off $E_{c} \approx 11$\,keV, spectral index $\delta \approx -6$, and power input $P \approx 10^{27}$\,erg\,s$^{-1}$, which we take as inputs to the model.  It is clear both observationally and from the modeling that a single loop model is insufficient, however, and in a multi-threaded model the power must be divided among many threads comprising an arcade.  \rhessi\ does not have the spatial resolution nor the cadence necessary to determine that distribution.  Since the HXR emission is cospatial to and strongly correlated with the intensity of TR lines (e.g. Paper \rom{1}, \citealt{cheng1981,poland1982,woodgate1983,poland1984,simoes2015b}), we use the intensity distribution of \ion{Si}{4} emission measured with \iris\ to estimate this distribution of energy carried by the electron beam to various threads.

In Paper \rom{1}, we presented a histogram of \iris\ 1400 \AA\ intensities at various times, that were fit to a power law.  The slope $\alpha$ was determined to be $\approx -1.6$ at most times, ranging from about $-1.0$ to $-2.5$.  We therefore assume that the intensity distribution is correlated with the energy flux carried by the electron beam on each thread, thus composing a power-law of beam fluxes.  Although the correlation between TR lines and HXR emission is well established at the large scale, both spatially and temporally, future HXR instrumentation ({\it e.g.} \foxsi, \citealt{krucker2014}) should verify that this remains true at smaller scales.  

We have run 37 simulations with HYDRAD, adopting electron beam heating with $E_{c} = 11$\,keV, $\delta = -6$ (the average values measured with \rhessi), and beam flux ranging from $10^{8} - 10^{11}$\,erg\,s$^{-1}$\,cm$^{-2}$.  We assume that the heating lasts for 10 seconds on each thread, with a flat temporal envelope.  We then calculate the \ion{Si}{4} and \ion{C}{2} emission from each simulation along the loop at all times, using the \iris\ response function.  

From this set of simulations, on a power-law distribution with slope $\alpha(t)$, varying in time as measured with \iris, we then randomly sample $N$ total threads to comprise the emission (threads can be used more than once).  We allow these threads to occur randomly at an average rate of 1 per $r$ unit time ({\it i.e.} on a Poisson distribution with average waiting time $r$), and assume $N \times r \gtrsim 600$\,s in order to last through the period under consideration (approximately 14:14-14:24 UT).  We then sum the emission from all $N$ threads to calculate a light-curve and line profile at all times as if they were all contained in the same \iris\ pixel, and then fit the profiles to calculate the Doppler velocity as a function of time.  We present the intensities and velocities at both 1 and 8 second integration times in order to compare directly with the observations (8 second integration) and with what might be seen by a faster instrument.  Since the cross-sectional area of an individual strand is unknown and necessary to calculate line intensities, we assume the pixel area is divided evenly among the $N$ strands, and additionally assume a filling factor of 1.  

\subsubsection{Synthesized light-curves and Doppler shifts}

There are many variables at play that cannot be directly measured, despite the abundance of instruments that observed the event: the number of threads comprising the emission $N$, the rate of heating bursts onto these threads $r$, the minimum and maximum sizes of energy release, $F_{\text{min}}$ and $F_{\text{max}}$ (if any).  We discuss ways to constrain these variables in this section.

\begin{figure}
  \begin{minipage}[b]{0.5\linewidth}
    \centering
    \includegraphics[width=3.3in]{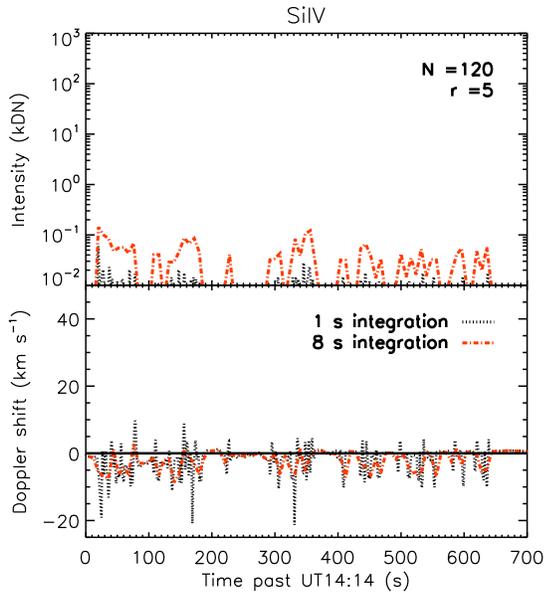}
  \end{minipage}
  \caption{Synthesized \ion{Si}{4} profiles for a set of randomly selected weak bursts, ranging in energy from $10^8 - 10^9$\,erg\,s$^{-1}$\,cm$^{-2}$.  The top plot shows the calculated intensity, and the bottom the Doppler shift.  Red-shifts are defined as positive velocities.  The line was calculated with integration times of one (black dotted) and eight seconds (red dashed), compared to the eight second integration time used with \iris\ in Paper \rom{1}.  Note the absence of significant red-shifts.}
  \label{fig:low_energy}
\end{figure}

First, consider the maximum size of the heating rate.  If that maximum is too small, the lines are generally blue-shifted, as in Figure \ref{fig:low_energy}.  We have limited the power-law to the range of energy flux $F_{0} = 10^8 - 10^9$\,erg\,s$^{-1}$\,cm$^{-2}$.  We show the emission calculated with $N = 120$ total threads and $r = 5$ seconds average waiting time.  The top plot shows the intensity of \ion{Si}{4} as a function of time, where the black dotted line shows a 1-second integration time and the red 8-second.  The bottom plot shows the calculated Doppler shifts, based on fitting a single Gaussian to the profiles, where we have defined red-shifted flows to be positive.  The line is, on average, weakly blue-shifted during the heating period ({\it cf.} \citealt{testa2014}), which was not observed during the event (see Paper \rom{1}).  It is clear that there must have been events with energy flux such that $F_{\text{max}} > 10^{9}$\,erg\,s$^{-1}$\,cm$^{-2}$.  

\begin{figure}
  \begin{minipage}[b]{0.5\linewidth}
    \centering
    \includegraphics[width=3.3in]{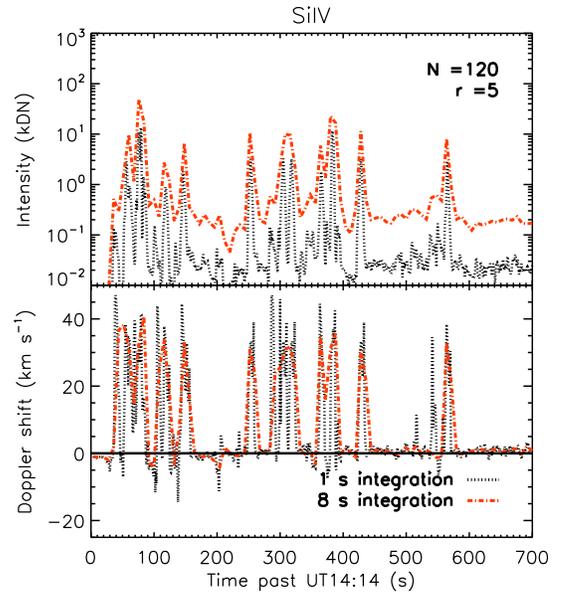}
  \end{minipage}
  \caption{Synthesized \ion{Si}{4} profiles for a set of randomly selected bursts, ranging in energy from $10^{8} - 10^{11}$\,erg\,s$^{-1}$\,cm$^{-2}$, otherwise as before.  The red-shifts are bursty and short-lived, in stark contrast to the observations.}
  \label{fig:all_energies}
\end{figure}

Figure \ref{fig:all_energies} shows the calculated light-curves in the case where the power-law extends from $10^{8} - 10^{11}$\,erg\,s$^{-1}$\,cm$^{-2}$ for $N = 120$ threads and $r = 5$\,s per thread.  At energy fluxes less than $10^{8}$\,erg\,s$^{-1}$\,cm$^{-2}$, there is little to no heating so that only minimal emission would result, while energy fluxes greater than $10^{11}$\,erg\,s$^{-1}$\,cm$^{-2}$ are so rare that we can discount them in general.  In this case, the stronger events quickly produce a red-shift in both lines as the chromosphere quickly heats to TR temperatures.  The stronger events also produce considerably higher intensities in both spectral lines, so that the emission is weighted heavily by them, despite being less common overall.  The net result is that the intensities are markedly more bursty in nature due to large events rising above the background of weak events, and that the calculated Doppler shift is weakly red-shifted in both lines when the intensity is high.  However, the velocities are not persistent over the period of heating, as in the observations.  There are two possible explanations: that the assumed minimum energy release is too low ($F_{\text{min}} > 10^{8}$\,erg\,s$^{-1}$\,cm$^{-2}$) or that the assumed spacing between events is too large ($r < 5$\,s per thread).  We examine both of these possibilities.  

\begin{figure*}
  \begin{minipage}[b]{0.33\linewidth}
    \centering
    \includegraphics[width=2.2in]{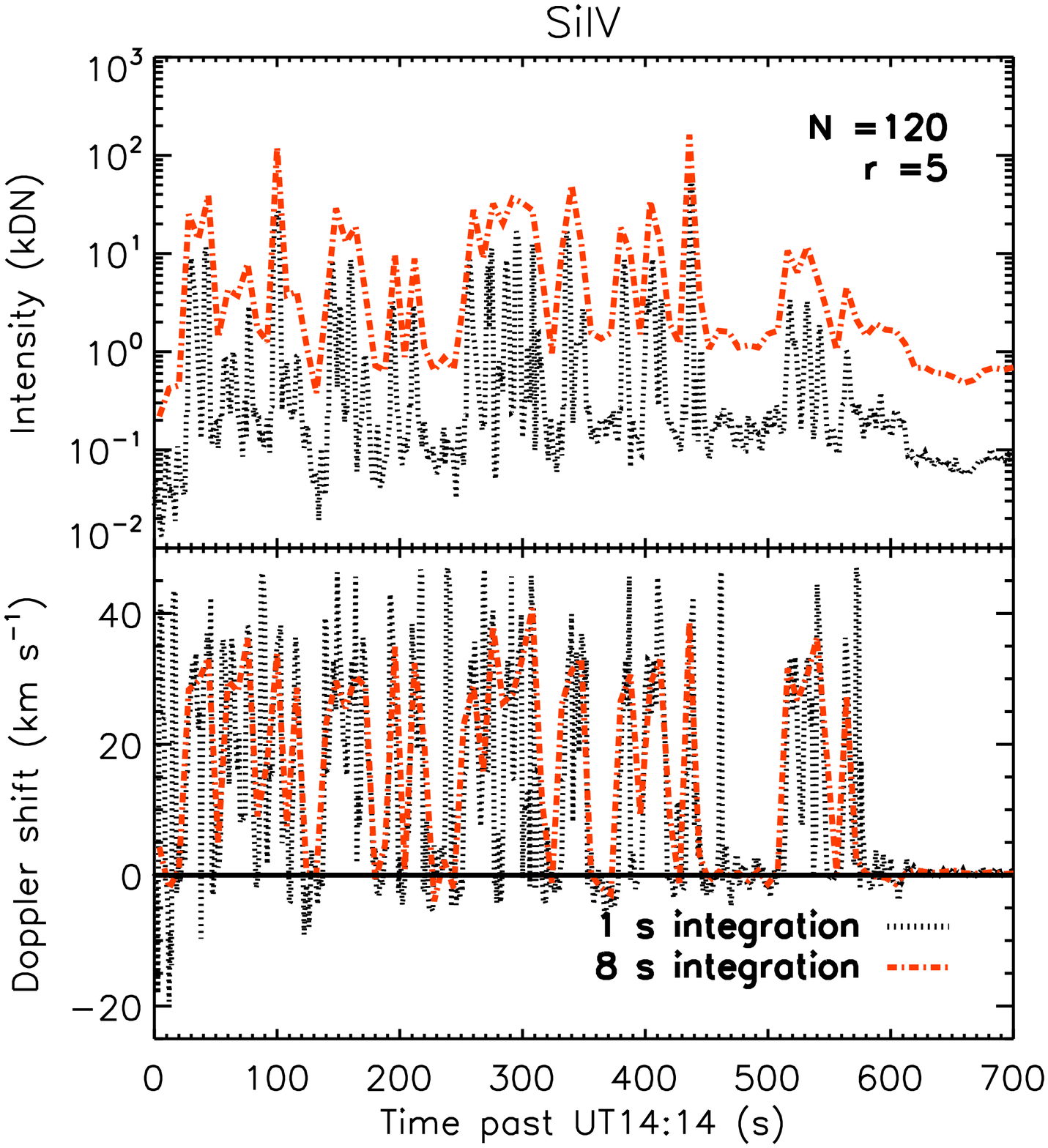}
  \end{minipage}
  \begin{minipage}[b]{0.33\linewidth}
    \centering
    \includegraphics[width=2.2in]{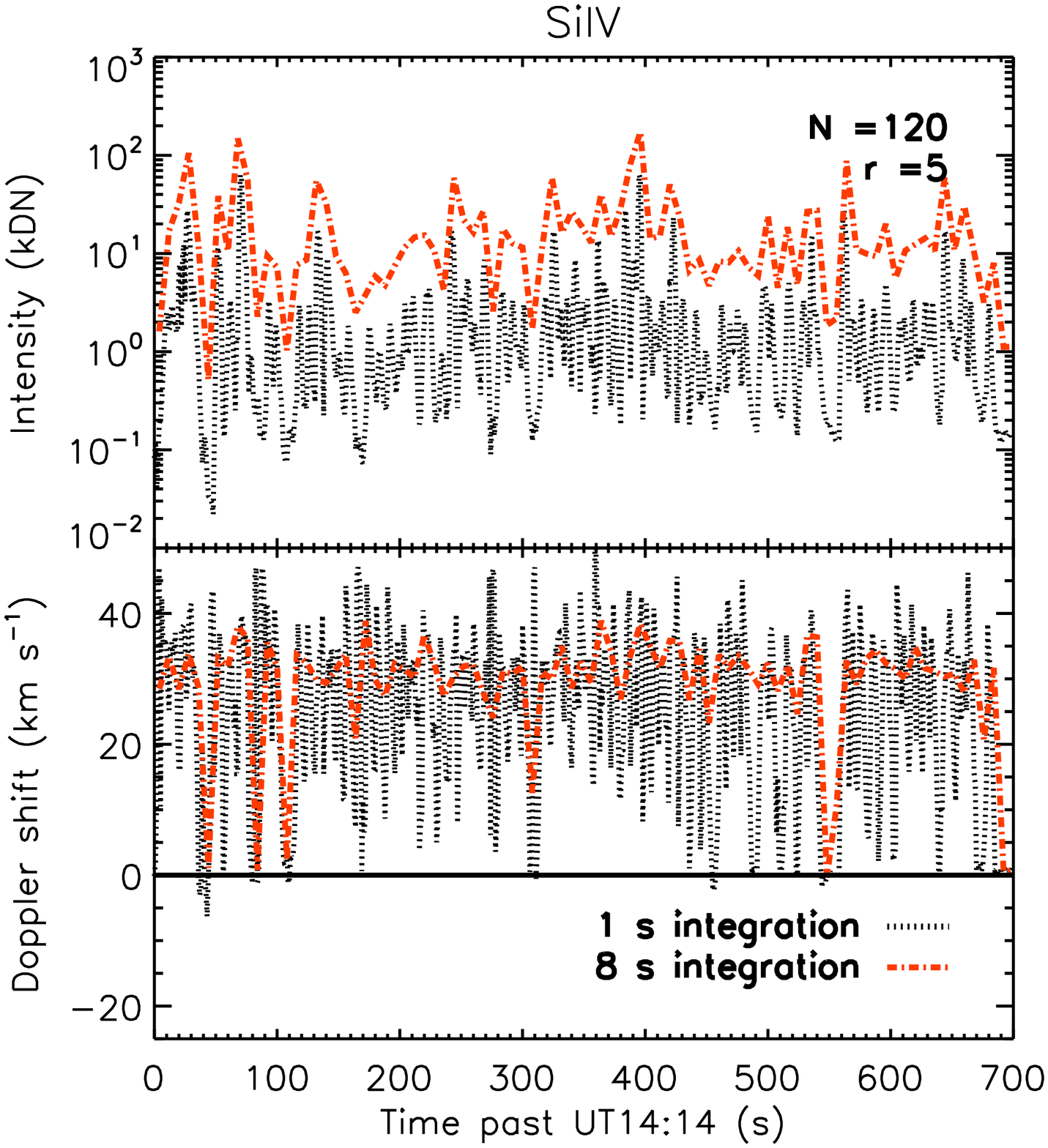}
  \end{minipage}
    \begin{minipage}[b]{0.33\linewidth}
    \centering
    \includegraphics[width=2.2in]{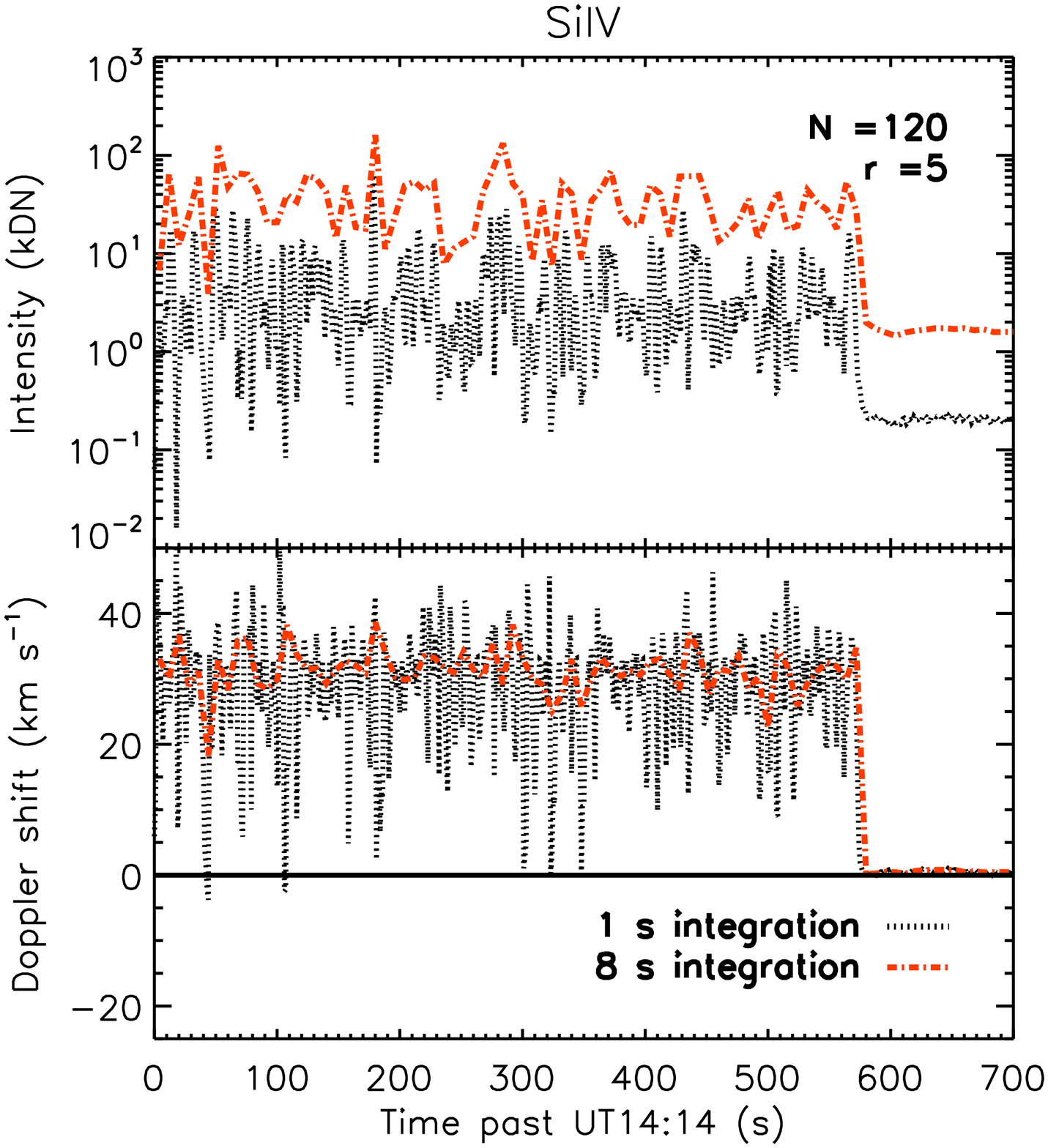}
  \end{minipage}
  \caption{Synthesized \ion{Si}{4} profiles for a set of randomly selected bursts, with minimum energy fluxes $10^{9}$, $3 \times 10^{9}$, and $5 \times 10^{9}$\,erg\,s$^{-1}$\,cm$^{-2}$, respectively.  The red-shifts become more persistent with larger minimum energy fluxes, and the sharp drops to zero disappear.  Interestingly, an instrument with a higher cadence might be able to detect drops in the velocity.  }
  \label{fig:minimum}
\end{figure*}

We now consider the minimum energy flux.  In Figure \ref{fig:minimum}, we show synthesized \ion{Si}{4} emission and Doppler shifts for the cases where the minimum energy flux is taken to be $10^{9}$, $3 \times 10^{9}$, and $5 \times 10^{9}$\,erg\,s$^{-1}$\,cm$^{-2}$, with $N = 120$ and $r = 5$\,s, which also can be compared with the previous figure.  For $F_{\text{min}} < 5 \times 10^{9}$\,erg\,s$^{-1}$\,cm$^{-2}$, there are sharp drops in the red-shift, which were not observed, while above that value, the red-shifts remain at around 30\,km\,s$^{-1}$ for the duration of the heating events.  This minimum beam flux corresponds roughly to the transition between gentle and explosive evaporation flows for the cut-off used here (see Section 5 of \citealt{reep2015}), suggesting that the majority of the threads were heated explosively.    

\begin{figure*}
  \begin{minipage}[b]{0.33\linewidth}
    \centering
    \includegraphics[width=2.2in]{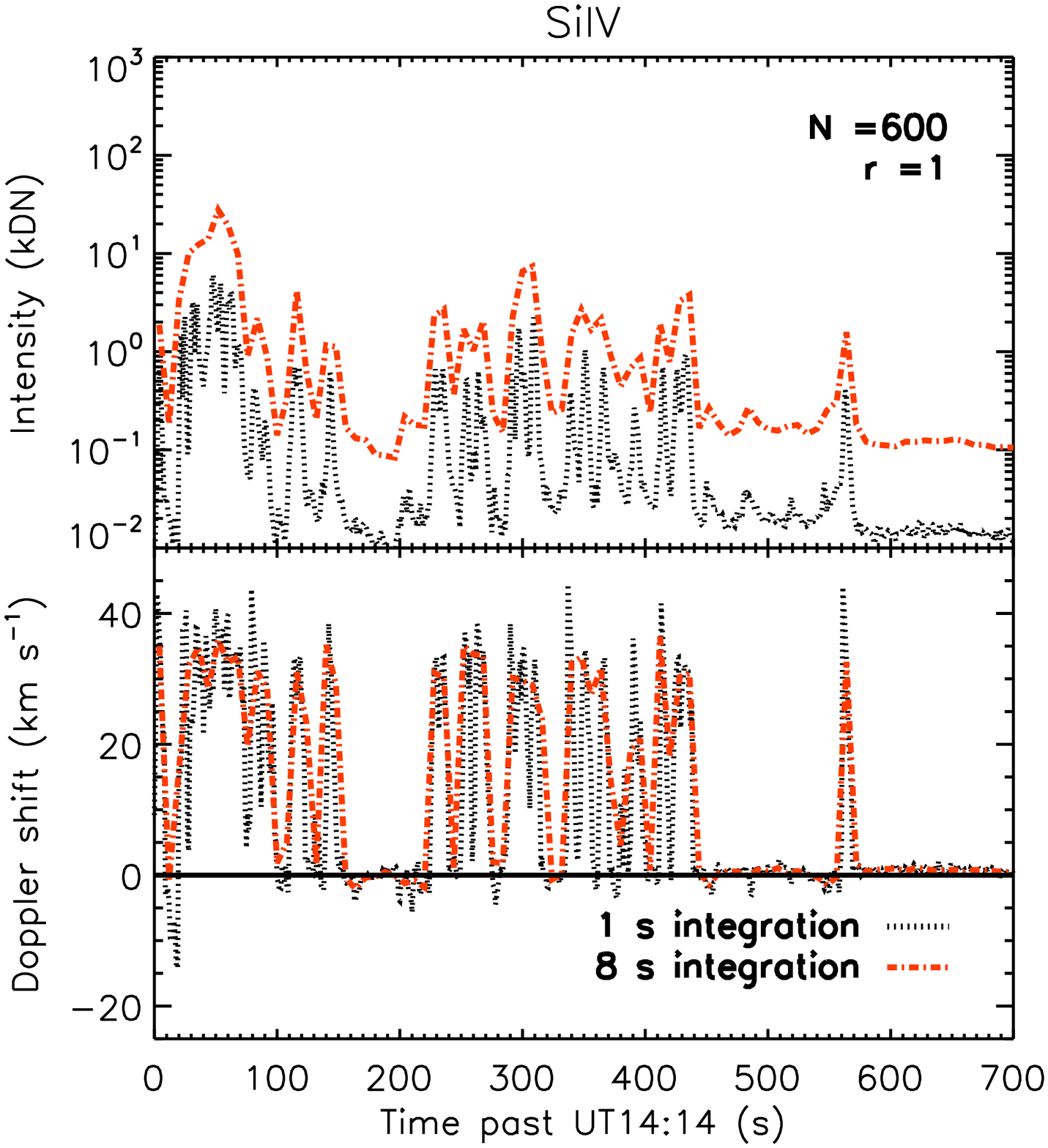}
  \end{minipage}
  \begin{minipage}[b]{0.33\linewidth}
    \centering
    \includegraphics[width=2.2in]{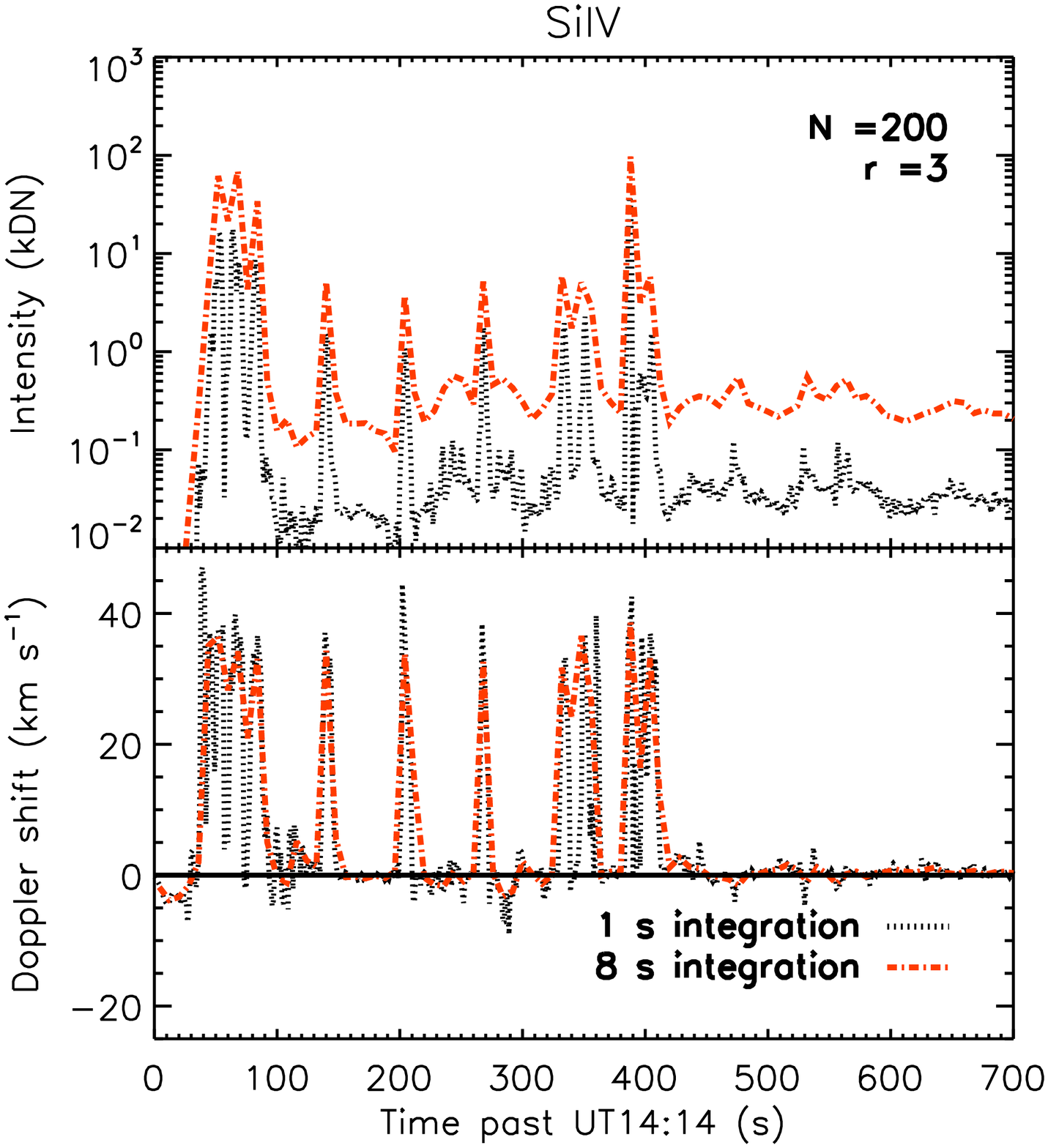}
  \end{minipage}
    \begin{minipage}[b]{0.33\linewidth}
    \centering
    \includegraphics[width=2.2in]{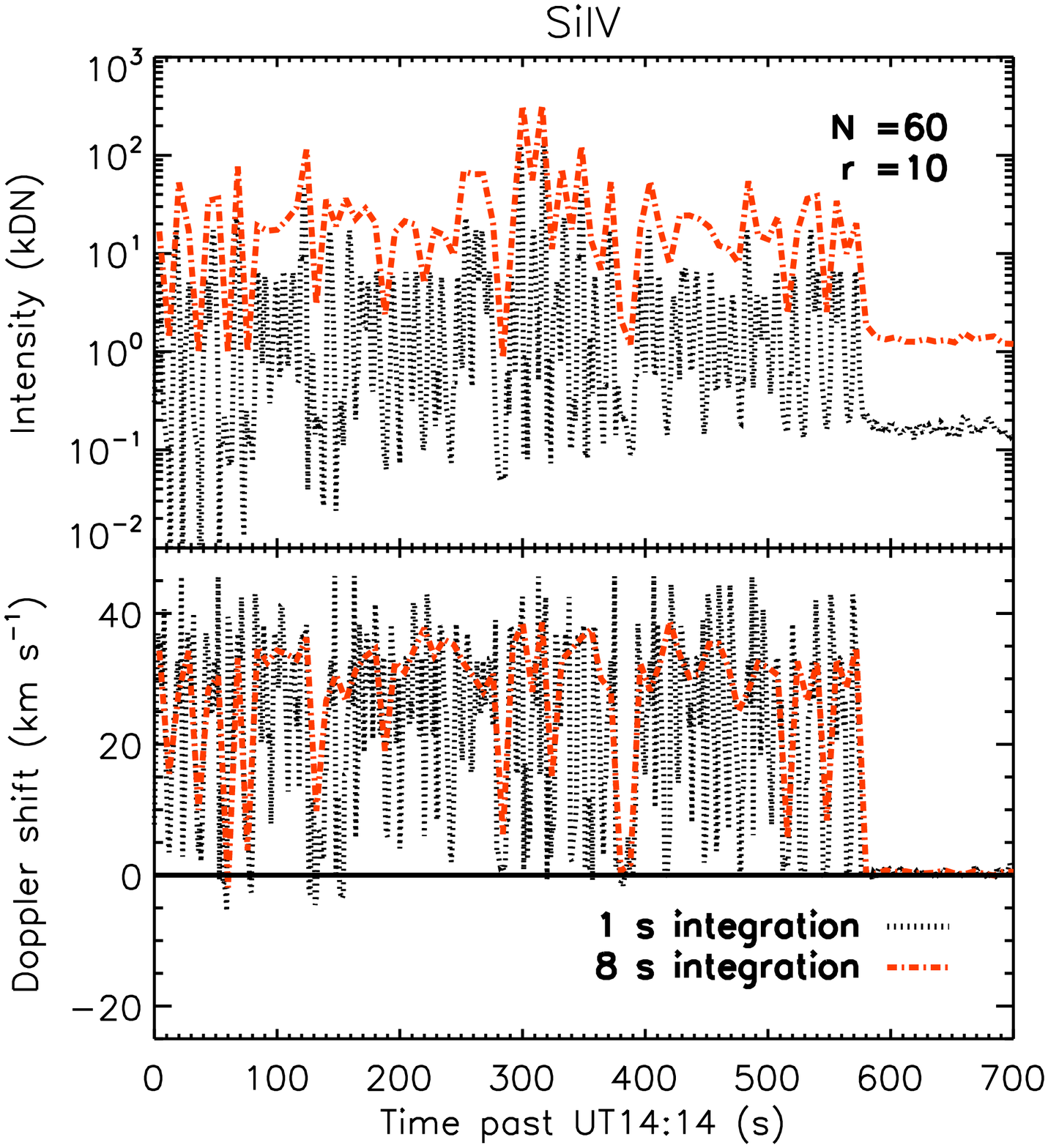}
  \end{minipage}
  \caption{Synthesized \ion{Si}{4} profiles, from left to right, with minimum energy fluxes $10^{8}$, $10^{8}$, and $5 \times 10^{9}$\,erg\,s$^{-1}$\,cm$^{-2}$, respectively, with $r$ values of 1, 3, and 10\,s per thread, respectively.  In all three cases, there are sharp drops in the red-shifts, indicating that the rate of new threads or minimum energy alone cannot explain the observed persistent red-shifts, but that both are required.}
  \label{fig:r}
\end{figure*}

One might ask whether the persistent red-shifts might also be explained by simply increasing the rapidity with which new threads are energized.  In Figure \ref{fig:r}, we show three plots, two where $F_{\text{min}} = 10^{8}$\,erg\,s$^{-1}$\,cm$^{-2}$, but now $r$ has been decreased to 1 and 3\,s per thread, and one where $F_{\text{min}} = 5 \times 10^{9}$\,erg\,s$^{-1}$\,cm$^{-2}$, with $r = 10$\,s per thread.  In all three of these cases, we see that the red-shifts have sharp drops toward 0, either due to the weighting of the weaker heating events or the quick decay of red-shifts on the individual threads.  However, we can deduce that: 1. $r < 10$\,s per thread, that is, new threads are energized at a rate faster than one per 10 seconds on average, 2. $N \times r > 600$\,s, so $N > 60$ threads must have been energized during the HXR burst, and 3. the rate of energization $r$ alone cannot explain persistent red-shifts, there must also be a minimum energy flux in general.  There is one caveat to mention, though, in that a constant low-energy cut-off was assumed on each thread, which cannot be verified with \rhessi, although the parameter strongly affects flow speeds \citep{reep2015}.

\subsubsection{Checks on the model}

The synthesized emission appears consistent with the observed values, given the constraints discussed above.  However, an important check on the model is to synthesize other observables to find if it is well bounded by them.  Can the model simultaneously reproduce the \iris\ velocities, the EM distribution found with \eis, \xrt, and \aia, and the density constraint found with \eis?  How do the line profiles compare to those found in Paper \rom{1}?

\begin{figure*}
  \centerline{\includegraphics[width=0.85\linewidth]{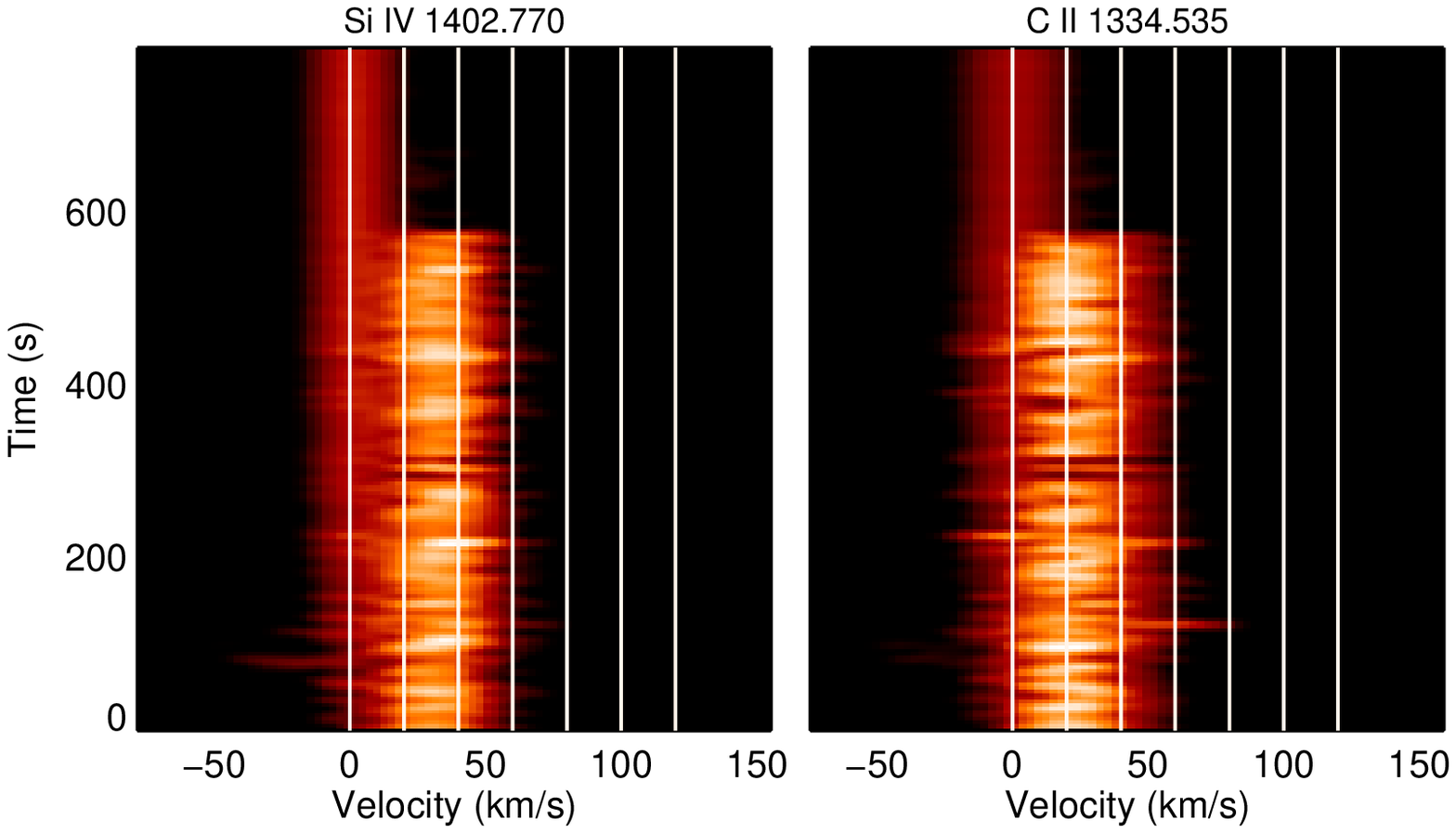}}
  \centerline{\includegraphics[width=0.5\linewidth]{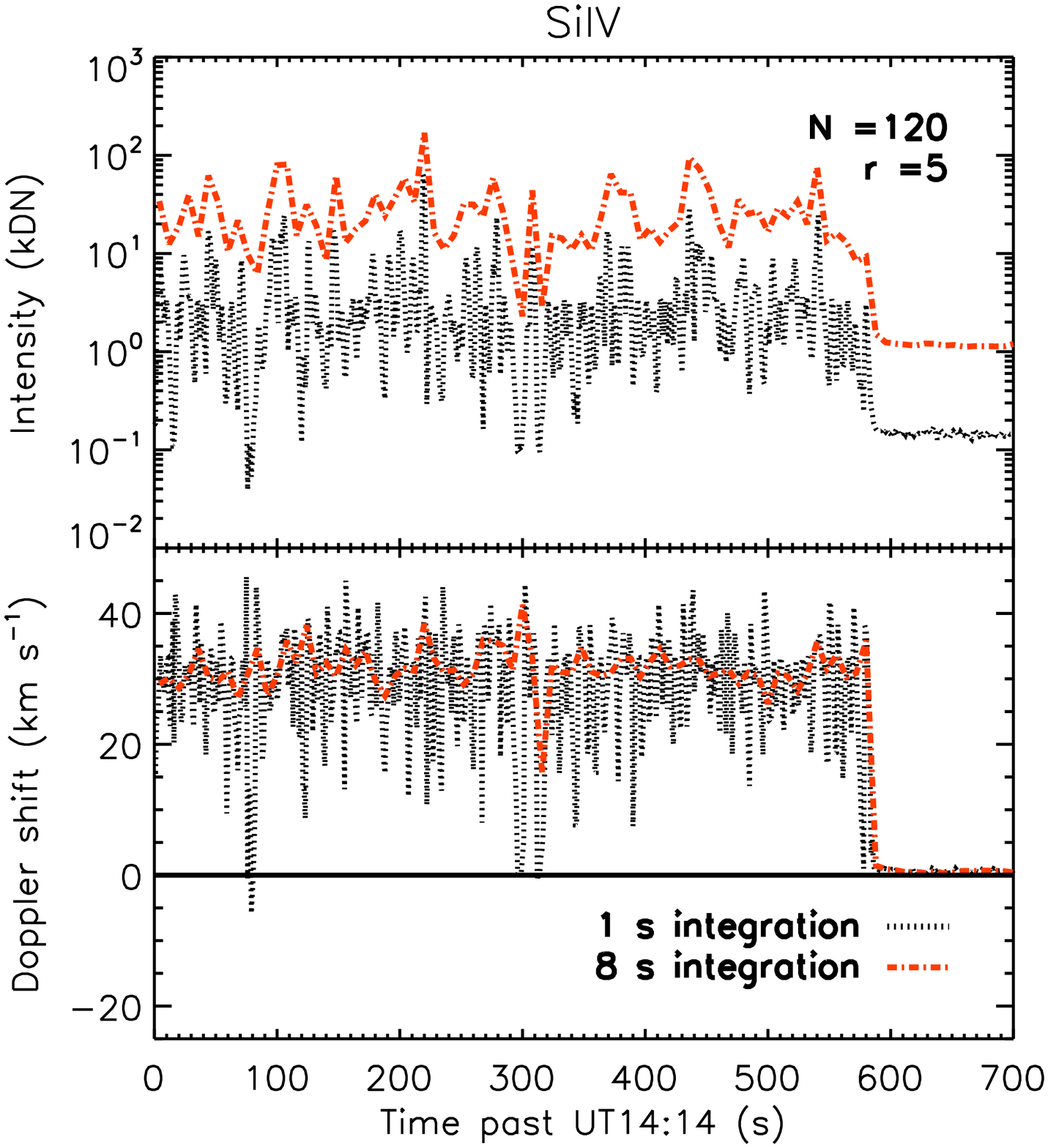}%
    \includegraphics[width=0.5\linewidth]{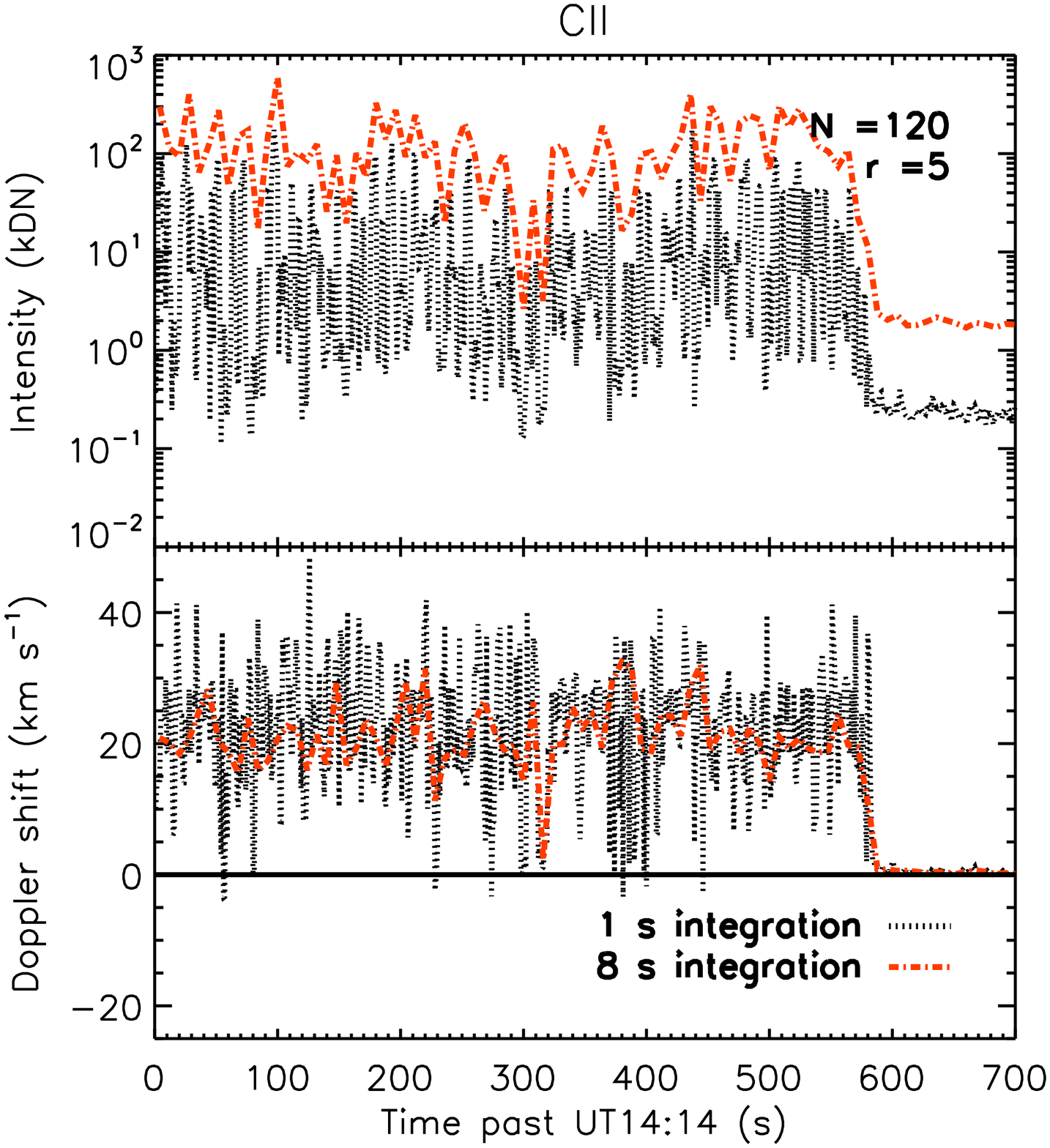}}
  \vspace{-0.2in}
  \centerline{\includegraphics[width=0.5\linewidth]{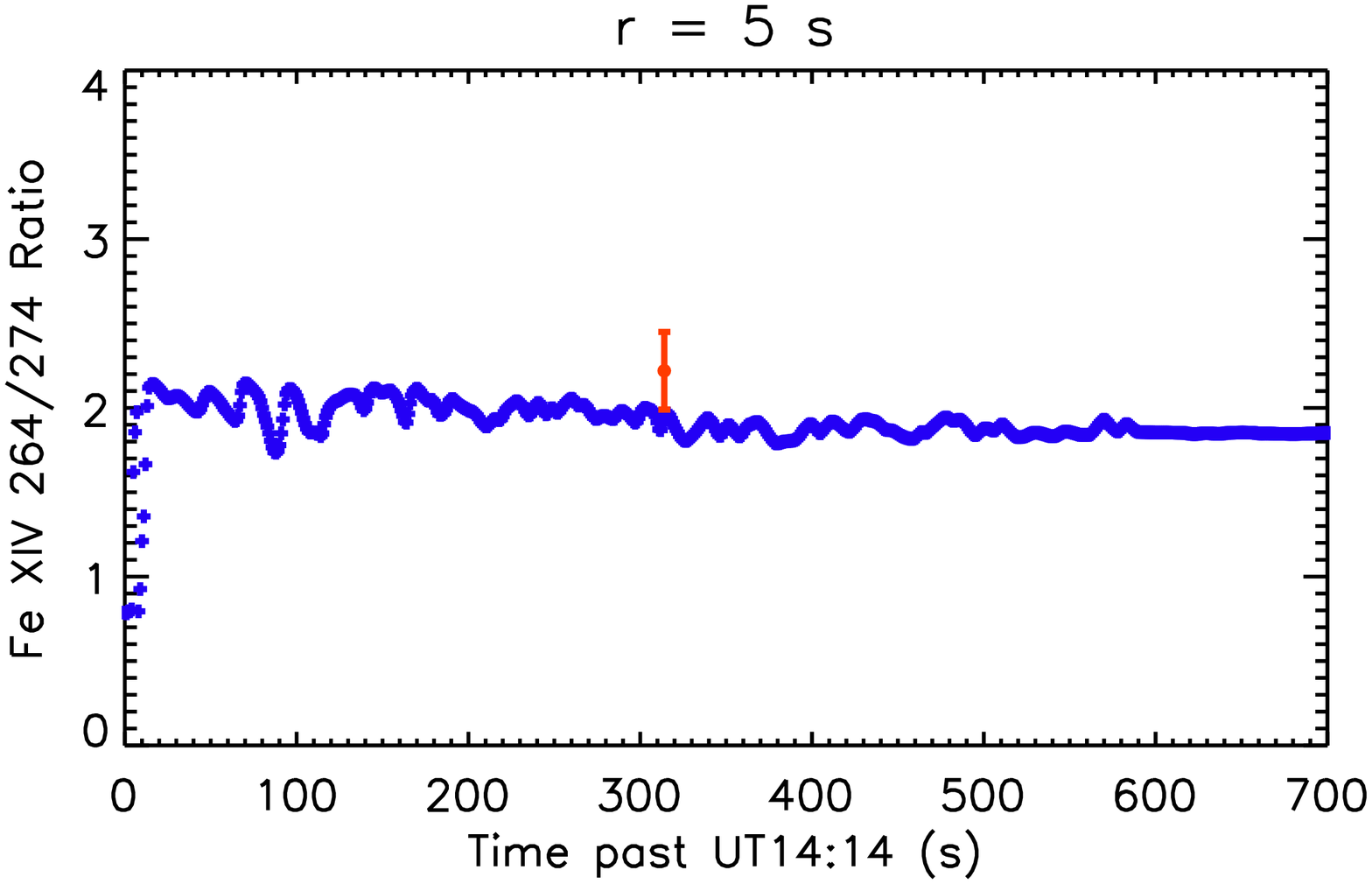}%
    \includegraphics[width=0.5\linewidth]{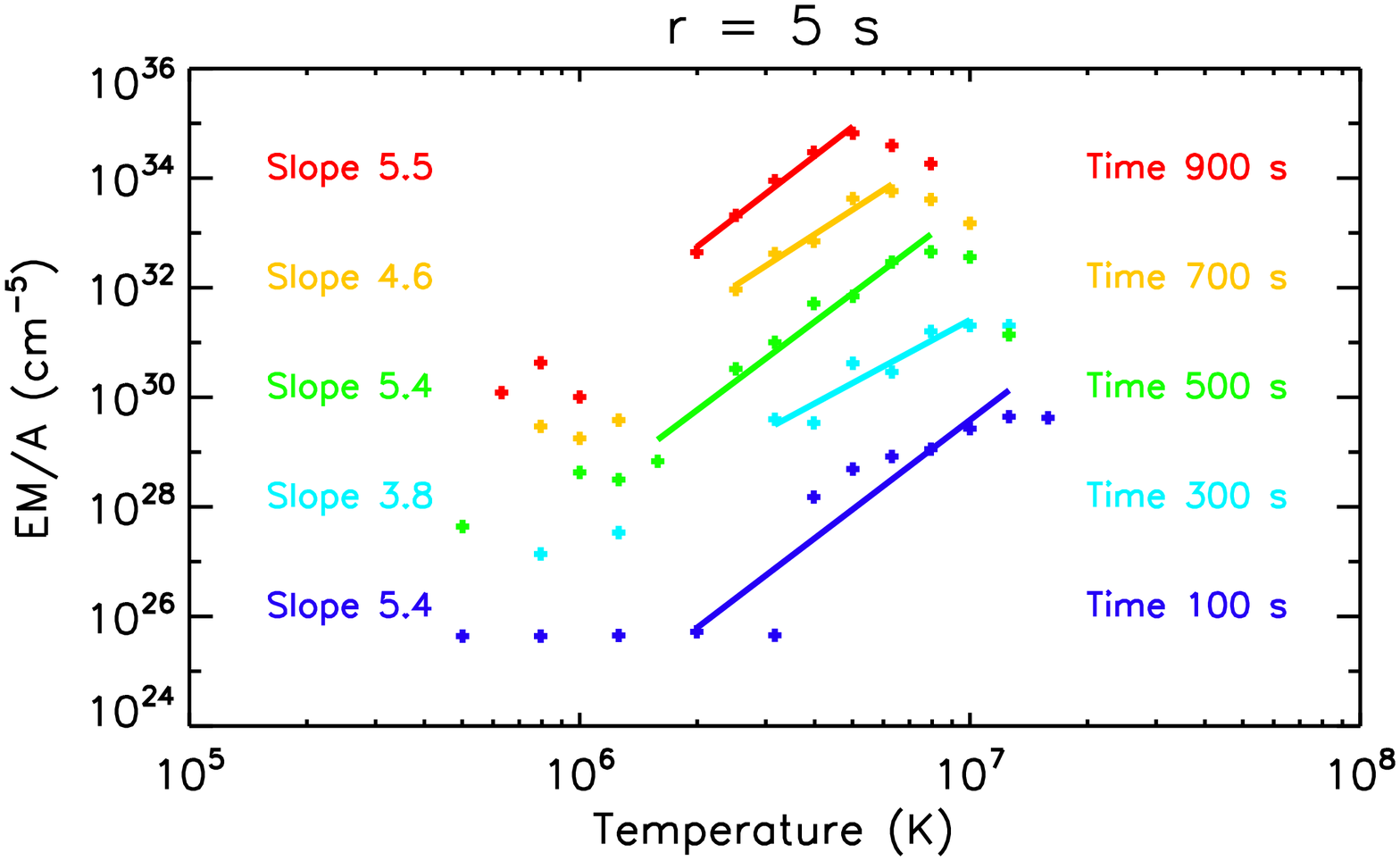}}
  \caption{$r = 5$\,s per thread, $N = 120$ threads.  At top, the line profiles for \ion{Si}{4} and
    \ion{C}{2} at 8-second integration, which can be compared with the observed values in Paper
    \rom{1}.  At center, the light-curves, as before.  The bottom left plot shows the ratio of the
    \ion{Fe}{14} $264.8$ and $274.2$ lines as might be seen by \eis, where the red data point
    indicates the observed value.  Finally, the bottom right plot shows the emission measure per
    area ($= n^{2} \Delta s$) integrated over 1\,arcsec near the apices of the loops, at a few
    selected times (multiplied by factors of 10 for clarity).  }
  \label{fig:other_params}
\end{figure*}

Consider the case where $r = 5$\,s per thread, $N = 120$ threads, $F_{\text{min}} = 5 \times 10^{9}$\,erg\,s$^{-1}$\,cm$^{-2}$, and $F_{\text{max}} = 10^{11}$\,erg\,s$^{-1}$\,cm$^{-2}$, which nicely reproduces the persistent red-shifts in both \ion{Si}{4} and \ion{C}{2}.  Figure \ref{fig:other_params} shows the line profiles and light-curves for \ion{Si}{4} and \ion{C}{2}, the emission measure per area (defined as $n^{2} \Delta s$, where we have integrated over 1 arcsec near the apices of the loops), and finally the density sensitive line ratio \ion{Fe}{14} $264.8/274.2$\,\AA\ as might be seen by \eis\ near the foot-point (the red data point denotes the observed ratio).  The observed line profiles in Paper \rom{1} can be compared with those calculated here.  Importantly, at most times, both lines show a bright red-shifted component, with a much weaker stationary component ($\lesssim 10\%$).  This compares favorably with the observed \ion{Si}{4} profiles, although the stationary component of \ion{C}{2} is smaller than that observed (Figure 9 of Paper \rom{1}).  The \ion{C}{2} line likely requires a full radiative transfer calculation to reproduce accurately ({\it e.g.} \citealt{lin2015}), as it forms in the upper chromosphere or base of the TR, whereas the contribution function in CHIANTI cuts off below $10^{4}$\,K.  The forward model only considers thermal broadening in determining line widths, and in general both spectral lines are thinner than the observed ones, suggesting that another broadening mechanism contributes to the observed values.  

The density measured from the \ion{Fe}{14} ratio is slightly lower than that observed, but within the error bars.  No significant trend was found between the number of threads $N$ and the calculated ratio, although higher minimum flux values increase the ratio, in general.  The line ratio oscillates more for smaller number of threads, although the cadence of the \eis\ raster is too slow to draw any conclusions from this.  

The emission measure distribution (EMD, e.g. \citealt{graham2013,simoes2015b}) reveals a good deal about the dynamics.  The maximum temperature in the loops tends to decrease with time - peaking at about 15, 10, 8, 6, and 5 MK at the times shown.  A line was fit to the cool-ward side of the EM from its peak value down to log $T = 6.2$, as done in Paper \rom{1}, with temperature bins of 0.1 d $\log{T}$.  The slopes of those lines steepen during the heating period gradually, since the hottest material dominates the emission during this time, and then become shallower as the loops cool after the heating ceases, as might be expected.  At 500 seconds, the calculated slope is about $5.4$ which is intermediate to the observed values measured with \aia\ and \xrt\, $6.4 \pm 0.9$ at UT14:22:49, and that found with \eis, $4.5 \pm 1.1$ (see Figure 13 of Paper \rom{1}).  

\begin{figure}
  \centerline{\includegraphics[width=\linewidth]{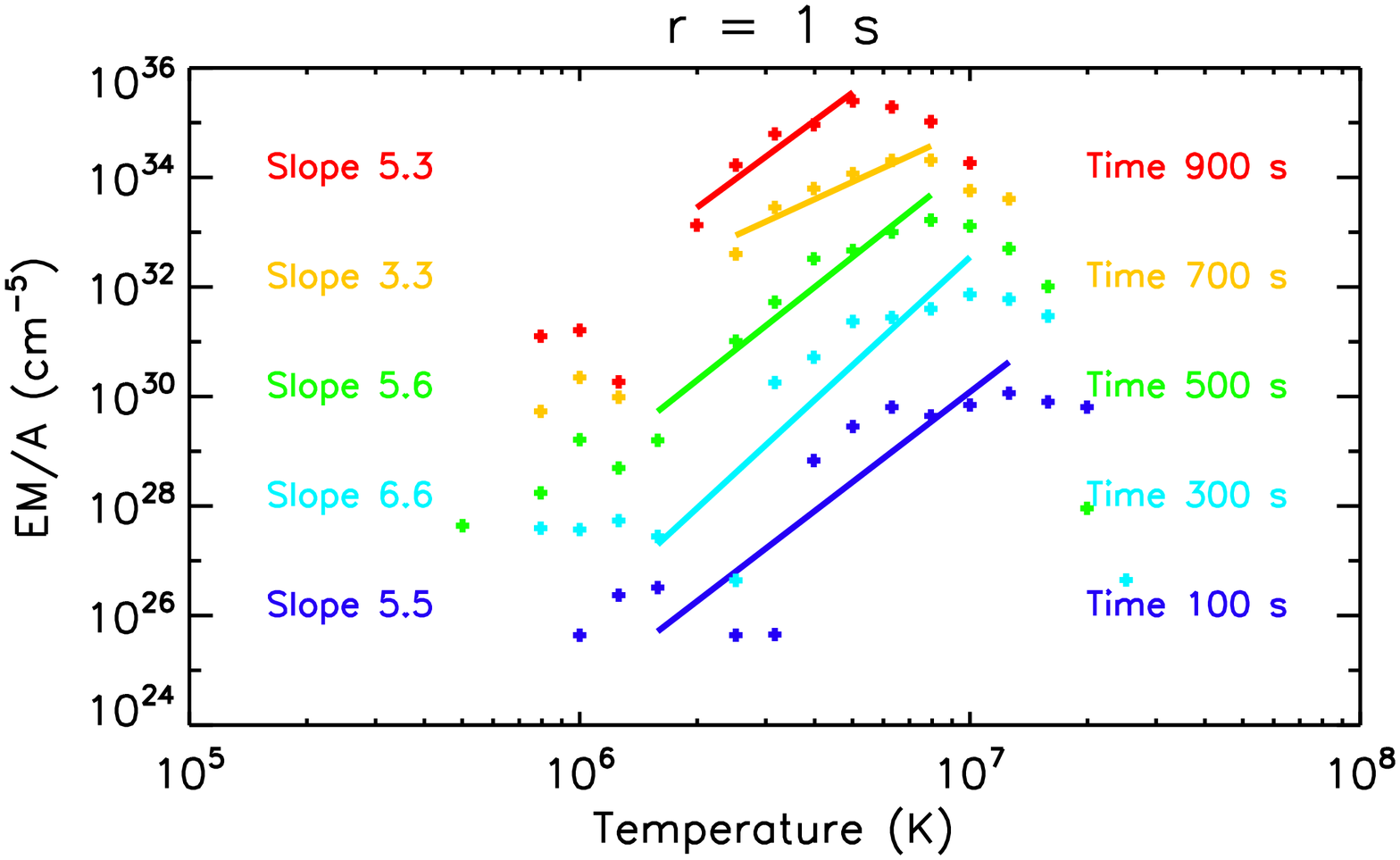}}
  \centerline{\includegraphics[width=\linewidth]{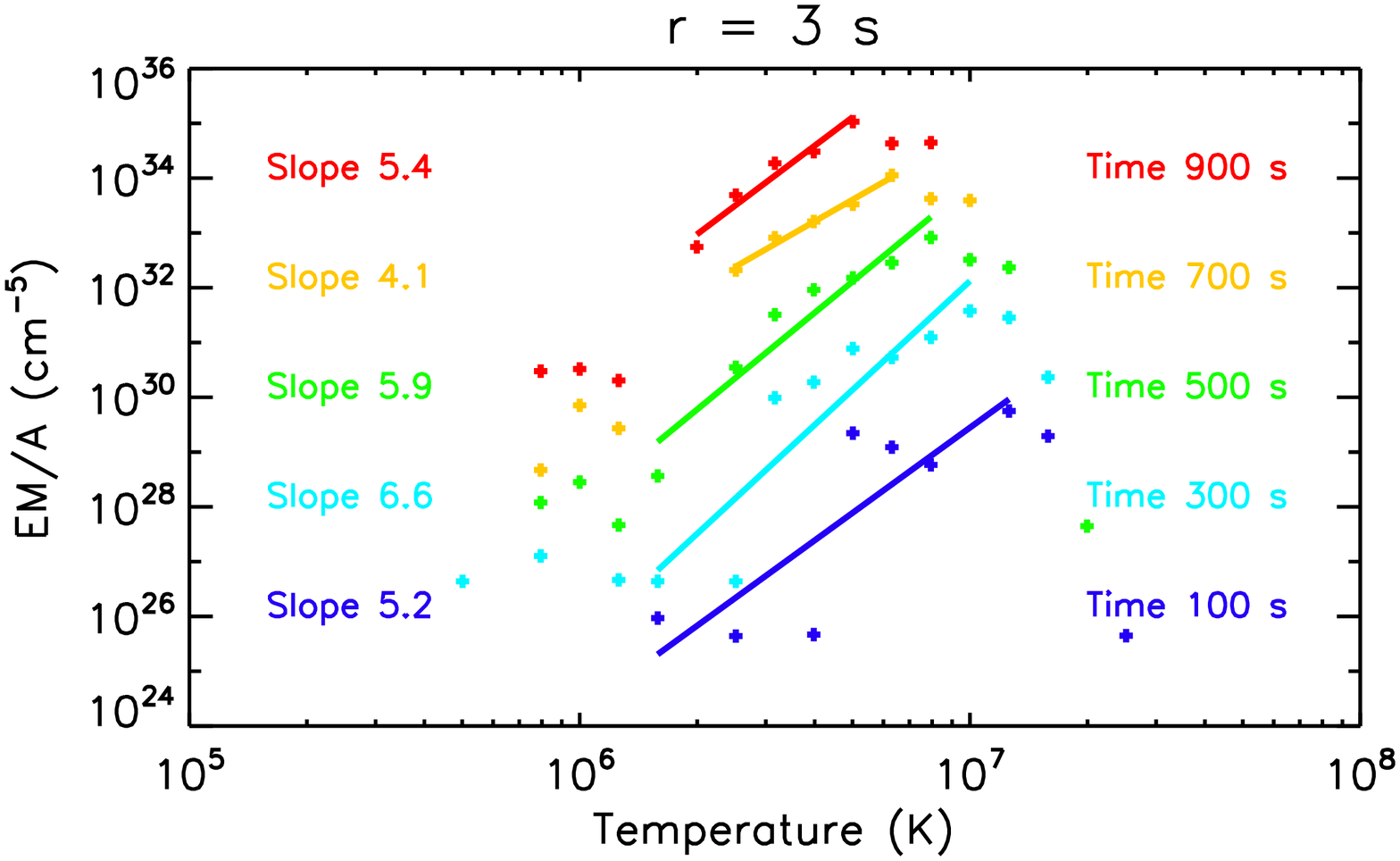}}
  \centerline{\includegraphics[width=\linewidth]{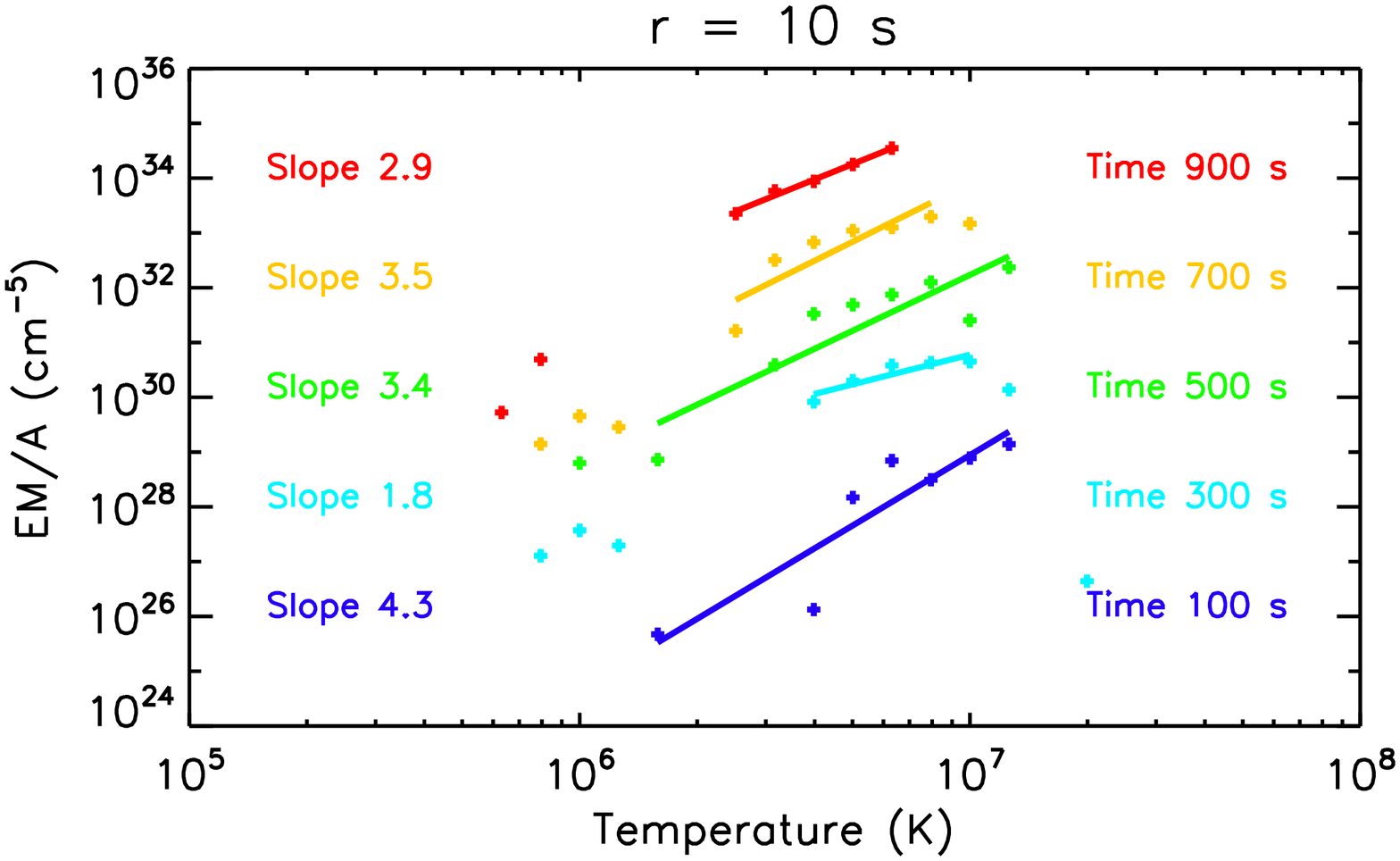}}
  \caption{Emission measure plots for $r = 1, 3, 10$\,s per thread, with $F_{\text{min}} = 5 \times
    10^{9}$\,erg\,s$^{-1}$\,cm$^{-2}$, and $F_{\text{max}} = 10^{11}$\,erg\,s$^{-1}$\,cm$^{-2}$.}
  \label{fig:em}
\end{figure}

In Figure \ref{fig:em}, we show the EMD calculated for $r = 1, 3, 10$\,s per thread, which can be compared with the previous EMD plot.  There is no clear trend in the slopes for varying values of $r$, despite what may be expected, which is likely due to the randomized energy and timing of individual threads.  However, for smaller values of $r$, where the number of threads $N$ is much larger, there is more emission at lower temperatures than in the case of {\it e.g.} $r = 10$\,s per thread, because there are many threads cooling at a given time.  In general, however, the slopes and peak temperatures are consistent with the observed values found in Paper \rom{1}.  

In general, we find that the model simultaneously reproduces the persistent red-shifts and intensities along with the EMD, maximum temperature, and density measured from \ion{Fe}{14}, given the numerous constraints mentioned above.  Most importantly: 

\begin{enumerate}
\item{The energy partition between threads is described by a power-law with slope $\approx -1.6$ as determined observationally}
\item{The average time between threads $r \lesssim 10$\,s per thread}
\item{The number of threads $N > 60$ within a single \iris\ pixel}
\item{A faster rate $r$ of new threads cannot explain the persistent red-shifts by itself.  There also must be a minimum heating flux on the majority of threads.}
\item{The results here suggest that for this event the minimum flux $F_{\text{min}} \gtrsim 3 \times 10^{9}$\,erg\,s$^{-1}$\,cm$^{-2}$, or more generally that the majority of threads are heated explosively. }
\end{enumerate}

\section{Implications \& Conclusions}
\label{sec:implications}

In Paper \rom{1}, we presented observations of the flare SOL2014-11-19T14:25UT, which was observed with many different instruments, covering a wide range of energies and temperatures.  Observed red-shifts in the \ion{Si}{4} 1402.770 \AA\ and \ion{C}{2} 1334.535 \AA\ lines persisted for longer than 30 minutes, which is difficult to reconcile with simple theoretical models.  Specifically, \citet{fisher1989} showed that condensation flows persist for about $45$ seconds, regardless of the strength or duration of heating.  

Alternatively, \citet{brosius2003} suggested that a ``warm rain'' scenario can produce long-lasting red-shifts in TR lines (see also \citealt{tian2015}).  During the impulsive phase of a large M-flare, \citet{brosius2003} found \ion{O}{3}, \ion{O}{5}, \ion{Mg}{10}, and \ion{Fe}{19} were all initially blue-shifted.  The two oxygen lines gradually transitioned into down-flows that lasted for half an hour, while the \ion{Mg}{10} line was found to be composed of a strong stationary component and a weaker red wing, and \ion{Fe}{19} remained stationary thereafter.  Those red-wing components, termed ``warm rain'', were interpreted as signatures of the cooling and draining of a loop, and lasted for half an hour or so after the flare's onset.  

However, the event studied here differs in a few important respects.  First, there were no signatures of blue-shifts in the TR lines during the impulsive phase.  \ion{Si}{4} was fully red-shifted for the duration of the event (Figure 9 of Paper \rom{1}), in contrast to the behavior of the oxygen lines reported in \citet{brosius2003}.  Second, there is insufficient time for the loops to drain.  The red-shifts begin simultaneously with the HXR burst, suggesting that they are signatures of chromospheric condensation as the energy is deposited by electron beams.  After heating ceases on a given coronal loop, there is a long time period during which the coronal density does not drain significantly, and energy losses are first dominated by thermal conduction, then by radiation, and only then by an enthalpy flux (see the thorough treatment by \citealt{bradshaw2010}).  The time scales for coronal loops to cool and drain were derived analytically and checked numerically by \citet{cargill1995,bradshaw2005,bradshaw2010,cargill2013}, and typically are on the order of 45 minutes to an hour.  Finally, the cooling between successive \aia\ channels often seen in flares ({\it e.g.} \citealt{petkaki2012}) was seen in the coronal section of the loops, with a cooling time-scale of about 40 minutes, suggesting they did not drain significantly for nearly as long.  

In this paper, by adopting a multi-threaded model, we have shown that these observations are consistent with a power-law distribution of heating occurring on a very large number of threads.  The following important conclusions can be drawn from the work here.

\begin{enumerate}
\item \textbf{Multi-stranded heating.}
The single loop model is woefully inadequate to explain the intensities or Doppler shifts observed in this event, regardless of the number of heating events on the loop or duration of heating.  A simple multi-stranded model of 7 loops similarly fails, as the observed Doppler shifts are essentially continuous, not single discrete events.  However, a multi-stranded loop model as presented in Section \ref{subsec:montecarlo} captures many of the observed properties of the \iris\ emission, while being within the bounds of the observed density, temperature, and emission measure.  Compare the conclusions of many prior multi-threaded studies, {\it e.g.} \citet{hori1998,reeves2002,warren2006}, etc. 
\item \textbf{Energy partition among strands.}
We measured the distribution of \iris\ SJI intensities to be well described by a power-law, with slope $\alpha = -1.6$ at most times (see Paper \rom{1} for details).  Since the intensities of many TR lines are proportional, both spatially and temporally, to the HXR intensities (Paper \rom{1}, \citealt{cheng1981,poland1984,simoes2015b}), and since the energy flux of the electrons is proportional to the non-thermal HXR intensity \citep{brown1971,holman2011}, we take this distribution as a proxy for the partition of energy among the threads.  This distribution over a large number of threads produces \iris\ \ion{Si}{4} and \ion{C}{2} intensities and Doppler shifts that are consistent with values measured in Paper \rom{1}.  In future work, we will examine this distribution for more events of varying \goes\ class to determine statistical trends and properties.
\item \textbf{Resolving loop structures.}
It does not seem possible to explain the observed red-shifts with a single loop model, or with a small number of strands.  Further, the background level of emission does not strongly show Doppler shifts (blue or red), and the shifts correspond to brightenings above background emission so that the red-shifts must be a signature of the flare itself.  As the red-shifts were measured in single pixels, then, we conclude that there is loop structure not being resolved at the sub-pixel level of \iris.  In order to maintain a red-shift in these lines without sharp drops in the speed, threads must be energized at a rate $r < 10$\,s per thread, giving a lower limit on the number of threads $N > 60$ {\it rooted within a single \iris\ pixel} for the duration of the HXR burst.  For the duration of the entire event, this number must be appropriately increased.  In comparison, \citet{simoes2015} estimated a rate of $r = 3$\,s per thread (total of 120 threads during the impulsive phase) over the entire reconnection region of a small C2.6 flare.  Their analysis was based on the released non-thermal energy, and constitutes a lower bound. \\
What is the size, then, of an individual strand?  If the IRIS pixel were divided evenly between strands, then the diameter is on the order of $\frac{1}{100}$\,arcsec or less, significantly smaller than previous suggestions.  This may provide evidence for the fractal model of reconnection in flares \citep{shibata2001,singh2015,shibata2016}, where the current sheet becomes exceedingly thin due to the secondary tearing instability \citep{zweibel1989}.
\item \textbf{Beam energy flux constraints.}
\rhessi\ measures the power contained in the electron beam integrated over the entire foot-point, which can then be divided by an area to give an estimate of the beam energy flux.  However, since it is integrated over the entire foot-point, that does not specify what the flux was on the many threads comprising that area.  Combined with the power-law distribution, we have constrained the maximum and minimum values of the flux.  The cut-off energy during this event was measured at 11-13 keV for the duration of the HXR burst (Figure \ref{fig:rhessi_params}).  At that cut-off, the threshold between gentle and explosive evaporation is $\approx 3 \times 10^{9}$\,erg\,s$^{-1}$\,cm$^{-2}$ \citep{reep2015}.  For lower beam fluxes, the \ion{Si}{4} line is in fact blue-shifted (compare \citealt{testa2014}), which was never observed during this event.  We therefore can reasonably conclude that the maximum beam flux must be greater than this value.  What's more, since small events are far more likely on a power-law distribution of energies, they strongly weight the emission and often cause sharp drops in the measured Doppler shift, so that it seems likely that the majority of the threads were heated explosively.  
\end{enumerate}

This work has given a great deal of insight into the dynamics of this small flare.  We have reasonably found a lower limit to the number of magnetic field threads, and have found the partition of energy among them, which allows us to build a realistic multi-threaded model.  This model is well constrained by the abundance of observations from many different instruments, and can be applied to flares which do not have coverage as good as this one.  There are still many areas of this work that can be improved to remove assumptions and generalize the model, however, such as determining how the electron beam parameters vary from thread to thread or finding an upper limit to the number of threads.  It is also often true that \ion{Si}{4} has a stronger stationary component in other flares than was seen in this one ({\it e.g.} \citealt{tian2014}), so that further work may be required to determine whence the difference arises.

We speculate that spectral lines seen in larger flares such as \ion{Fe}{21} 1354.08\,\AA\ may further improve our understanding of energy deposition between threads.  The results of \citet{fisher1989} make it clear that the duration of condensation flows are insensitive to the heating strength and duration.  However, evaporation flows are not limited in the same manner, and in fact there are indications that the flows last as long as the heating does ({\it e.g.} the flows in Figures 4 and 5 of \citealt{reep2015} or the \ion{Fe}{21} shifts in \citealt{polito2016}).  Unfortunately, for this event, \ion{Fe}{21} was not observed, and so no hard conclusions can yet be drawn regarding the heating durations. 


\acknowledgments The authors thank Lucas Tarr, George Doschek, and the anonymous referee for discussions which improved the modeling and understanding of this event.  This research was performed while JWR held an NRC Research Associateship award at the US Naval Research Laboratory with the support of NASA.  The research leading to these results has received funding from the European Community's Seventh Framework Programme (FP7/2007-2013) under grant agreement no. 606862 (F-CHROMA) (PJAS).  IRIS is a NASA Small Explorer developed and operated by LMSAL with mission operations executed at NASA Ames Research center and major contributions to downlink communications funded by the Norwegian Space Center through an ESA PRODEX contract.  CHIANTI is a collaborative project involving George Mason University, the University of Michigan (USA) and the University of Cambridge (UK).


%

\end{document}